# Use of Data Mining in Scheduler Optimization

George Anderson, Tshilidzi Marwala, and Fulufhelo V. Nelwamondo

School of Electrical Engineering, University of Johannesburg, South Africa

Email: georgeganderson@gmail.com,tmarwala@uj.ac.za,fnelwamondo@csir.co.za

**Abstract**: The operating system's role in a computer system is to manage the various resources. One of these resources is the CPU (Central Processing Unit). It is managed by a component of the operating system called the CPU scheduler. Schedulers are optimized for typical workloads expected to run on the platform. However, a single scheduler may not be appropriate for all workloads. That is, a scheduler may schedule a workload such that the completion time is minimized, but when another type of workload is run on the platform, scheduling and therefore completion time will not be optimal; a different scheduling algorithm, or a different set of parameters, may work better. Several approaches to solving this problem have been proposed. The objective of this survey is to summarize the approaches based on data mining, which are available in the literature. In addition to solutions that can be directly utilized for solving this problem, we are interested in data mining research in related areas that have potential for use in operating system scheduling. We also explain general technical issues involved in scheduling in modern computers, including parallel scheduling issues related to multi-core CPUs. We propose a taxonomy that classifies the scheduling approaches we discuss into different categories.

1. INTRODUCTION

Operating Systems consist of several components such as the scheduler, memory manager, file system manager, and I/O device manager. These components have several parameters which are used to tune them to suit a particular workload. A workload consists of units of resource utilization, such as processes and threads, which will make use of the various resources in the computer through the appropriate resource managers (scheduler, memory manager, etc.) [1, 2, 3].

In this survey, we consider one of these resource managers: the CPU (Central Processing Unit) scheduler. Schedulers, for example the one in the Linux operating system kernel, have various parameters that are set with little in the way of theoretical foundations backing these settings (Gandhi, 2007). Operating system performance parameters, including scheduler parameters, are set in an iterative cycle of tuning and measuring [4, 5]. This involves measuring components of the system such as CPU utilization, memory usage, and throughput. This iterative process consumes a lot of time [6, 7, 8]. It would be of great benefit to system engineers and administrators to have a system built into the operating system kernel that will do away with the need for a system administrator to tune certain parts of the operating system. Apart from tuning schedulers, other scheduling problems exist. Some of these problems are related to scheduling on multi-core architectures. Modern CPU designs have shifted from trying to increase clock speeds of CPUs to incorporating more processor cores on a single CPU chip, providing facilities for parallel execution of programs, similar to the traditional SMP (Symmetric Multiprocessing) architectures where a computer has several CPUs [9]. Instead of having multiple CPUs, modern computers have multiple cores on a single CPU, which is cheaper and more energy efficient. This presents challenges in scheduling. For example, when two threads competing for resources are scheduled to run at the same time, on different cores for example, performance suffers. However, when two communicating threads are not scheduled to run concurrently (i.e. are not co-scheduled), performance also suffers [10].

Data mining is currently a popular tool for data analysis. Because of progress in data acquisition, storage, and other Information Technology tools [11], huge databases are being created. In order to avoid organizations becoming data rich and information poor, data mining was developed to extract information from such huge databases [12]. In this paper we survey the operating system scheduling problem and discuss various solutions to it that are based on data mining. We also look at data mining solutions to related problems that have potential impact on operating system scheduling. To the best of our knowledge, there is no published research that gives an overview of such solutions to scheduling in one place. These tools are very powerful and a good fit to the scheduling problem, since the process of scheduling generates a huge amount of data which can then be analyzed for decision making, for example how to schedule processes and threads. This research paper is organized as follows: Section II gives an overview of software performance optimization methodology; Section III discusses scheduling in the operating system; Section IV gives an overview of data mining; Section V is the main section in this paper, giving an overview of scheduling solutions based on data mining; Section VI summarizes recent trends discussed earlier in the paper and Section VII concludes.

## 2. SOFTWARE PERFORMANCE OPTIMIZATION

Computer system performance is closely related to the time it takes for certain processes to be completed and is part of Quality of Service (QoS). The less time it takes to complete these processes, the better the performance. Some of the metrics used in performance include response time, throughput and scalability. Response time is measured from the time a user of a computer system submits a request to the time a result is produced e.g. a user clicks a button on a word processor and it carries out an action such as changing the font of an entire document. The lower the response time value, the better the performance is deemed to be. This is particularly important in interactive systems [2]. Throughput is the number of processes that complete per unit time. Scalability is also important. A system is said to be scalable if its performance degrades gracefully (i.e. not significantly) as the load increases [13].

Other important metrics, especially as related to scheduling, include average waiting time (i.e. how long a process spends waiting for the CPU on average), average turnaround time (i.e. average time it takes for a process to complete its work), and resource utilization (i.e. the percentage of time a resource such as the CPU is actually executing a process' instructions). A resource utilization value of 95% is very high and in a server environment this would suggest that the scheduler is doing a good job of allocating the CPU to processes. Various approaches are used to study the performance of a system [13]:

- Experimentation: In this approach a system, perhaps a prototype is built, and its performance is measured. Decisions are then made as to which aspects need to be modified. This approach is very accurate but also very expensive.
- Simulation: A computer model of a system is constructed and executed. This model captures only the relevant aspect of a system. Both static and dynamic properties are represented in the simulation model. Various settings can be played around with to test various ideas. This approach is accurate and could provide a very high level of detail but the simulation may take a long time to develop, validate and run, so it is not good, for example, for self-adaptive systems, which have to analyze their environment and change their behavior on-the-fly.
- Analytical Modeling: A model is developed using formulas and/or computational algorithms. Systems of equations are solved to give various values used to predict performance of a computer system. This approach is very fast to execute (to solve the equations) but they may take time to develop as good mathematical skills are necessary, and may not be as accurate as the other two approaches.

In this paper, we survey various approaches to scheduler optimization based on data mining. These approaches make use of experimentation and analytical modeling. In the next section we look at issues related to CPU scheduling in operating systems.

## 3. SCHEDULING IN THE OPERATING SYSTEM

A computer has various hardware components such as the CPU (Central Processing Unit), main memory, hard disk drives, printers, keyboard, mouse, a monitor, and network interfaces. Application programmers write their programs to run on specific types of computers. In order for each application programmer not to be concerned with how to interact with the various hardware components or how to manage these resources, modern computers have a layer of software around the hardware called the operating system [3, 1, 2]. The job of the operating system is to manage the various resources in the computer, to make sure they work well together. The reason why users use a computer is not because of the operating system, but rather because of the application software. This software is applied to tasks in the real world e.g. word processing, spreadsheets, and playing games. However, this software does not work on its own because of the reasons given earlier. Thus we have the system software. This category of software includes the operating system, and other programs such as file managers and anti-virus software. The job of system software is to manage and maintain the computer system and support the application software.

The operating system is always running, from the moment the computer has finished starting up to the moment it is switched off, all the time doing its job: managing resources. Its performance therefore has a big role in the performance of the overall computer system. Each application program running in a computer is represented by the operating system using a process.

The processes compete for CPU time because that is how they are executed i.e. they are carrying out their computing when they are making use of the CPU. It is the job of the CPU scheduler to allocate CPU time to processes [1, 3]. Schedulers use scheduling algorithms to carry out this allocation. Certain algorithms are more appropriate for certain types of workloads (collections of processes executing on a computer system) than for others. Examples of basic scheduling algorithms, i.e. those found in introductory operating system text books; include First Come First Served (FCFS), Shortest Job First (SJF) and Round Robin (RR). Variations of these have been developed for commercial operating systems. A particular scheduling algorithm may have one or more parameters that can be adjusted to tune the CPU allocation process e.g. the time quantum (how big a continuous time block a process has with the CPU), and various priority levels.

Faxen et al. [9] wrote a report on the state of the art of CPU designs. Current CPU designs have multicore architectures, also known as Chip Multi-Processors (CMP). They consist of several cores which function like individual CPUs. Each core has a fast Level 1 cache, and can share a larger Level 2 cache with the other cores, as in the Intel Core 2 Duo CPUs. Some designs have each core with its own Level 1 and Level 2 cache, and sharing a Level 3 cache with the other cores, as in the AMD Phenom and Intel Core i7 processors. Also, a Level 2 cache could be shared by a subset of the CPU cores, as in the Intel Core 2 Quad CPU. The caches store data and instructions for access faster than fetching them from main memory, thereby feeding the fast cores quickly. Schedulers typically implement time sharing, where each thread gets a small time slice on a processor or core, and space sharing, where each job is assigned to a subset of the cores available. A group of cores could be sharing a level in the cache hierarchy. Threads can then be scheduled on cores in the same group, for communication efficiency between threads, or spread over several groups, maximizing aggregate cache size to minimize cache misses. It is important for the operating system to use methods such as constructive cache sharing for threads to find the data they need in the cache as a result of the previously executing thread using the same data. Modern CPUs have a variety of performance counters that can give figures on events such as cache misses, thus the operating system could work out which threads needs are not being met by their cache allocations.

Frachtenburg and Shwiegelshohn [10] describe the current challenges facing the parallel job scheduling community. For commodity parallel computers such as desktops and servers they identify several scheduling challenges, including variable loads (interactive applications, media applications such as video, background programs, and parallel applications that require synchronization or co-scheduling). The underlying problem is that co-scheduled processes suffer degraded performance, while collaborating processes suffer performance loss if not co-scheduled. For desktops the scheduler must also respond to changes in user priorities quickly.

Kumar [14] describes the internals of the Linux CFS (Completely Fair Scheduler). The Linux CFS scheduler keeps track of how much tasks are being treated unfairly. A task is being treated unfairly if it is not being executed by the CPU. It maintains a binary search tree (a red-black tree) that orders tasks according to how unfairly they are being treated. A task that has been treated the most unfairly (because it has been waiting to use the CPU for a long time) is selected to execute on the CPU at the next opportunity. The CFS scheduler is modular, making it readily extensible, and also supports group scheduling and scheduling classes. It was introduced in Linux kernel version 2.6.23. The kernel also featured a modular scheduler, whereby schedulers can be selected at boot time, to suit various workloads (note the distinction between modularity for extensibility and modularity for replacement of scheduling algorithms). Jones [15] describes Linux kernel support for Symmetric Multiprocessing (SMP). He explains that the major mechanisms for SMP support involve ensuring load balancing across processors or CPU cores and processor affinity, ensuring threads and processes do not move around from core to core very much, thus being able to access data and instructions from the cache. Dow [16] also discusses the importance of processor affinity.

### 4. DATA MINING AND SYSTEMS

Giudici [17] defines data mining as: the process of selection, exploration and modeling of large databases in order to discover models and patterns that are unknown a priori." Data mining refers to the whole process of data extraction and analysis, which produces decision rules for specified business goals; data mining has many uses in business

[18]. For this study, we are concerned with production of decision rules for performance modeling in order to enhance the CPU scheduler. Data mining is different from data retrieval in that in data retrieval, the criteria for extraction are decided beforehand, so they are external to the extraction itself. In data mining, the relations and associations between data are not known beforehand but are discovered by the algorithms used [17, 19]. This is why data mining is sometimes called Knowledge Discovery in Databases (KDD) [20]. Data mining is also different from regular application of statistical procedures, which usually make use of experimental data. Data mining usually uses observational data.

Performance engineers can have access to large amounts of trace data, which documents various events taking place in the computer system, e.g. in the operating system. This data has the potential to predict the evolution of interesting variables or trends in a computer system but usually goes untapped or under-utilized. Modern software and hardware are cheaper and more powerful than before, allowing system engineers to collect and organize trace data in databases that give easy access to the data. Research in computing and statistics has produced flexible and scalable procedures that can be used to analyze such data stores.

Data mining methods come from the fields of machine learning and statistics [17]. Data mining methods that do not require a probabilistic formulation are from machine learning and are known as computational methods. Data mining methods that require a probabilistic formulation are from the field of statistics and are known as statistical models. Solomonoff [21] wrote the first paper on machine learning. He discusses inductive inference (the discovery of knowledge from observations). The work of Shapiro [22] is a very good example of the early use of machine learning in the area of systems. He made use of inductive inference in the design of a debugger that detects bugs by matching inputs with expected outputs. When the output is incorrect, it means there is a bug. The program (or piece of program) that caused this is then modified automatically to produce a similar program that generates correct outputs.

Liu et al. [23] developed data mining algorithms for detecting locations of logical bugs in programs. The programs in question are tested with appropriate data, which causes certain behaviors in logical statements. Where the behavior in a program differs significantly in a logical statement, for the case where the output is correct and where the output is incorrect, it suggests that that is the buggy section of code. Hirsh [24] gives an overview of where data mining was in 2008 and its future. He mentions various challenges, including data mining streaming data. The data is not stored anywhere prior to mining but is mined as it appears in a system. Another challenge in data mining discussed by Hirsh is how to take advantage of relatively new technology, such as multi-core CPUs.

Some future trends in data mining that could be useful for operating systems include capability to analyze change in trends over time and better user interfaces with reduced parameterization [20]. Gaber et al. [25] give an overview of mining data streams. According to them, areas that have been addressed include the handling of the continuous flow of data streams, unbounded memory requirements, and result accuracy. Open issues include developing interactive data mining environments, data pre-processing, use of new data stream mining technology, and real-time accuracy evaluation.

Operating systems can be modified to generate traces, which are records of their execution [26]. Various patterns can be recorded which could be used for debugging, security enforcement, performance optimization, and dynamic reconfiguration. However there is a lack of tools needed to analyze kernel traces [27]. Some of the problems that need to be solved in order to produce useful tools include mining large traces and semantic interpretation of results, which are both major problems in data mining [28, 29]. The next section gives a classification and overview of scheduling solutions based on data mining, which addresses some of the issues raised this section.

5. DATA MINING SCHEDULING SOLUTIONS

In this section we give an overview of various groups of scheduling solutions based on data mining concepts. We first propose taxonomy, and then discuss some representative solutions for distributed systems, single processor systems, symmetric multiprocessing systems, and asymmetric multiprocessing systems.

5.1. Data Mining Scheduling Solution Taxonomy

CPU scheduling solutions based on data mining can be grouped into those for distributed and centralized (non-distributed) systems. Silberschatz et al. define a distributed system as a collection of processors that do not share memory or a clock [1]. Such systems include clusters of computers, offering greater parallel computing power than a single machine could. Our survey is focused on centralized systems, but some approaches for scheduling distributed systems could be modified for centralized systems. Centralized systems could be those with a single processor or multiple processors/multiple cores. Multiprocessing systems could be symmetric, if all processing elements have equal capability and performance, or asymmetric, if the processing elements are not equivalent with one another. Figure 1 shows the taxonomy.

### 5.2. Scheduling of Distributed Systems

Zhang [30] implemented a system to aid resource scheduling in a distributed environment using machine learning techniques that are part of data mining. Her system trained by matching performance attributes of application behavior with classes. A new workload is then analyzed and classified automatically. This is then used to autonomously (i.e. without human intervention) schedule applications in virtual machines, for example, deciding that a CPU-intensive and an I/O-intensive program should run on the same machine, since they are using different resources, thereby minimizing competition for resources; Zhang's machine learning system runs on machines separate from machines running the target workload. Zhang focused on sequential applications running in virtual machines in distributed environments (e.g. clusters).

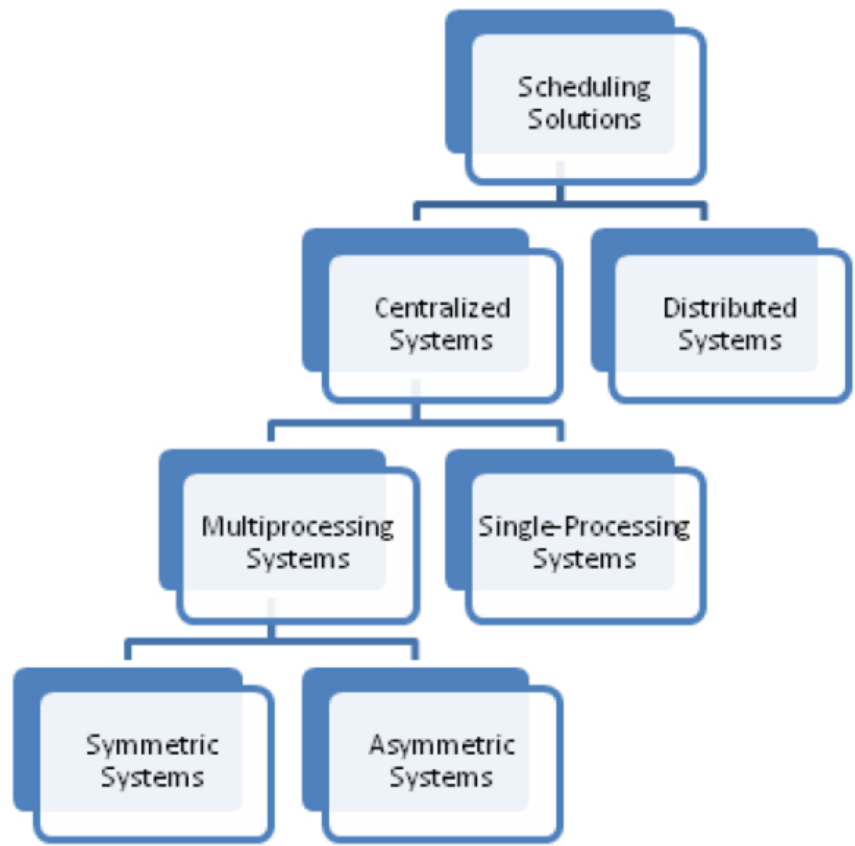

FIGURE 1. Taxonomy of Scheduling Solutions Focused on Non-Distributed Systems

### 5.3. Single Processor Systems

Xian [31] developed a system that involves interaction between a high level language execution environment (a Java virtual machine) and the host operating system. He developed two virtual machine-operating system schedulers that work together to schedule Java threads in such a way as to minimize the effects of garbage collection and synchronization between threads (lock contention, in particular). His system requires knowledge of the inner-workings of virtual machines. In his dissertation, Feng [32] discusses the development of a configurable scheduler, which approximates several scheduling algorithms by adjusting a single parameter online. His work confirms the importance of online tuning schedulers. The confirmation is done through the implementation of a new scheduler.

Nkgau and Anderson [29] carried out a study on using data mining on operating system scheduling traces for the purpose of discovering patterns which would be useful in trace compression, synthetic trace generation (to drive trace-driven simulators), and analysis of scheduling algorithms in operating systems. They made use of a commercial DBMS (Database Management System) to apply sequence clustering and association rules to kernel trace data. Their goal was to develop a trace-driven simulator to test scheduling algorithms. In their work, they performed off-line data mining. In the course of their work, they did not solve problems directly related to scheduling workloads but their framework supports this purpose.

Sopitkamol and Menasce [33] studied performance optimization of an e-commerce web site. They studied the effect of variation of various configuration parameters on performance in order to discover which parameters are more important for certain workloads. They used a statistical test to determine the largest impact parameters: ANOVA. Their work could be used in a system that tunes an operating system scheduler automatically, as part of a data mining system, where behavior of the effect of scheduling parameters on various workloads will be monitored and analyzed, followed by tuning of the scheduler.

LaRosa et al. [27] studied the application of data mining to kernel traces for the purpose of finding patterns which could indicate system problems such as bugs. They made use of a frequent item set algorithm which worked by splitting up a kernel trace into windows and then finding the number of windows in which events occurred. LaRosa et al. are continuing with their study by looking into building a library of normal execution patterns so that these could be eliminated from the patterns resulting from data mining, in order to eliminate noise. They are also looking into closely integrating their data mining framework with the kernel to produce an on-the-fly data mining capability as well as trying out other data mining algorithms. Their work focuses on finding bugs, however it could be used to tune an operating system scheduler, by analyzing traces online and adjusting the behavior of the scheduler appropriately.

The Computer Science Department of the University of Illinois at Urbana-Champaign has a project called SysMiner [34]. They are involved in System Mining, the application of data mining techniques to improve system performance, dependability and manageability, with a focus on Storage Systems.

One particularly interesting category of data mining algorithms for mining operating system trace data is frequent pattern mining, as used in the work of LaRosa et al. [27], mentioned above. There are various sub-categories including sequential pattern mining, structured pattern mining, correlation mining, associative classification, frequent pattern-based clustering, as itemized by Han et al. [28]. The survey done by Han et al. identifies frequent pattern mining as very useful in capturing the underlying semantics of data. For example, they mention the relevance to indexing, for example of event logs, which tries to reduce the amount of data to be scanned. Frequent pattern mining is also useful in mining data streams. This could be used to mine operating system traces as they are generated, on the fly, enabling the creation of a powerful tool to aid system engineers and administrators Han et al. also include software bug mining and system caching as important applications of frequent pattern mining. These are examples of system mining [34]. In their study, Han et al. give examples such as pre-fetching and disk scheduling as important applications. Although disk scheduling is also scheduling, it is different from the CPU scheduling we are interested in. For example, the service time given to a process has more restrictions than in CPU scheduling. Han et al. discuss open problems in frequent pattern mining. These include scalable mining methods (mining large operating system traces, for example, is a problem), use of approximate frequent patterns, powerful semantic interpretations of mining results, and use in new areas of research.

A study with significant implications for CPU scheduling, but in a different area is the work done by Zhang and Bhargava [6]. They developed a system built into an operating system kernel that uses machine learning to recognize workloads, select appropriate disk scheduling policies, and tune the selected disk scheduler. They tried to find out if such automation can improve efficiency and accuracy and how much performance it incurs. Five machine learning algorithms were used in their study: C4.5 Decision Tree, Logistic Regression, Nave Bayes, Neural Networks, and Support Vector Machine (SVM). Zhang and Bhargavas results indicate that their automation approach works. We are more interested in the CPU scheduler, not the disk scheduler. The data to be analyzed is therefore related to usage of the CPU, not disk accesses. Another relevant study is the work done by Moilanen and Williams [7] and also Moilanen and Schopp [35]. They developed a genetic algorithm library within an operating system kernel. Genetic algorithms are used to tune various parts of the kernel, such as the I/O scheduler and CPU scheduler. Various CPU scheduler settings are used and these evolve using genetic algorithms based on performance measurements. Their study yielded results that show their approach works. Moilanen and Williams suggest keeping a database of workload fingerprints to speed up analysis of workloads.

In their study, Teller and Seelam [8] provide insights into dynamic adaptation of operating system policies. They stress the importance of customizing the environment for certain types of applications using dynamic adaptations. They discuss the need for workload characterization and prediction techniques and how adaptations targeted at an entire system are more difficult than those targeted at a single application. They discuss various measures of the success of adaptations such as throughput and execution time and that other measures such as fairness, response time and latency may also be important. They identified several operating system components that may be suitable for dynamic adaption. These are: process scheduling (the focus of our study), I/O scheduling (for example, disk scheduling, as implemented by Zhang and Bhargava [6]), and memory management (for example, multiple page size management and virtual memory management). Other possible targets are network stacks, file I/O and scheduling of chip multiprocessors i.e. multi-core CPU scheduling, which our study also addresses. Teller and Seelam also recommend quantifying potential gains with static adaptation after selecting some workloads and benchmarks. This should be done in order to gain insights into potential performance improvements. Teller and Seelams work provides guidance on how to adapt operating system kernels.

Other work that provides guidelines on dynamic adaptation/self-tuning of operating system kernels includes that of Barrera [5]. He notes the importance of expectations of system performance, measurements of system performance, analysis of measurement data and actions to be carried out as a result of analysis, such as gathering more data and performing dynamic reconfiguration. Seltzer and Small [36] also provide guidelines for self-monitoring and self-adapting operating systems. They discuss the use of online and offline analysis of operating system data in order to adapt policies and parameter setting of operating system modules and their benefits. Benefits include making the system easier to manage, since it will be self tuning, and improving performance.

Reiner and Pinkerton [37] developed guidelines for adaptive performance of operating systems and conducted some experiments. They used statistical tests to recognize correlations between operating system performance and various scheduling parameter settings. Their results suggest that such dynamic adaptation yields performance gains.

Osisek [38] studied the use of machine learning in deterministic scheduling problems. Machine learning was used to learn how certain workloads could be decomposed (broken up into shorter scheduling sequences), based on previous requests. Once decomposed, his system would find out the best way to assign jobs to machines. Osiseks work attempts to leverage machine learning to improve scheduling. His work deals with deterministic scheduling problems, where processing times as well as other parameters are known beforehand. This could be suitable for a factory production line. For processes running on top of an operating system, such information is not known beforehand.

Li [39] studied the application of data mining to scheduling jobs on a single machine in a production environment. He used genetic algorithms to analyze historic scheduling records to capture the best performing schedules with the most important attributes and decision trees to enable creation of schedules for new workloads (jobs). As with Osiseks work, the focus was on deterministic scheduling, where various parameters are known beforehand, which is more suitable for a factory production line than an operating system.

### 5.4. Symmetric Multi-Processing and Multi-Core Systems

Fedorova et al. [40] studied manipulating thread priorities within a scheduler. They presented a solution to a problem with CMPs, cache-fair scheduling: a thread with a particular priority is affected by another thread (called a co-runner) with access to its cache; because of this, a high priority thread may not make expected progress because it is frequently waiting for cache misses to be resolved. Their solution involves boosting the time slices of threads so affected. Their implementation goes through a sampling phase where various parameters about a thread are collected, and a scheduling phase, where the threads time slice is adjusted. Sampling is done at regular intervals. One limitation of their approach is that phase changes in behavior of a thread are not detected.

Knauerhase et al. [41] developed a system to aid scheduling of tasks in operating systems on multicore architecture in a way that minimizes cache interference, ensures migration from core to core is done with minimum impact on performance, with knowledge of cache usage, ensures fairness to tasks, and accommodates functional asymmetry. Calandrino and Anderson [42] developed a real-time scheduler in the Linux kernel well suited for multicore architectures. They tackled two major problems of such a scheduler: automatic cache profiling and implementation efficiency. They made use of CPU performance counters to influence scheduling decisions.

Bulpin and Pratt [43] developed scheduling algorithms for the Linux operating system that are aware of Hyper-Threaded (HT) architectures i.e. Symmetric Multithreading (SMT). Their algorithm calculates the number of events per cycle and matches this with an estimate called performance ratio. These figures are obtained during training. During a live run, linear regression analysis is used to estimate the performance ratio for each process, and process priorities are raised to get them to run together with other processes in order to maximize the speed-up.

Chen et al. [44] studied two scheduling algorithms that could be used in CMP environments for constructive cache sharing. Work Stealing involves the scheduler using a queue for each core, where tasks are added at the top of the queue, but can be removed from either end. When a new thread is created, it is added at the top of the queue. The scheduler on a particular core selects a task for execution located at the top of the queue; this ensures related task contents are still in the cache. If the scheduler cannot find a task, it removes one from the bottom of another cores queue, thus ensuring the related tasks for the other core will execute together. The Parallel Depth First (PDF) algorithm selects tasks for execution that correspond with what a sequential program would do. An assumption is made that important sequential programs have been optimized to use the cache efficiently. Their study found that the PDF algorithm performs as well as, or outperforms, the Work Stealing algorithm.

### 5.5. Asymmetric Multi-Processing and Multi-Core Systems

There is widespread acknowledgement of the need for single-ISA (Instruction Set Architecture) performance asymmetric multicore CPUs [45]. Such CPUs implement the same instruction set but differ in performance characteristics such as cache size or clock frequency. Such CPUs achieve a desirable balance between performance, die area, and power consumption, compared to homogenous processors. Sondag and Rajan [45] developed an approach for assigning threads to asymmetric cores based on the observed IPC (Instructions per Cycle). Programs are instrumented statically with markers demarcating blocks of similar instructions. Similar blocks are grouped together. The blocks are observed when the program is running, on all cores. Experiments were carried out on a quad-core CPU, with two cores under-clocked i.e. slower. An assignment that maximizes the IPC of a thread running on a core is selected. A thread running on a fast core resulting in relatively few instructions per cycle is not desired. Such a thread should be moved to a slower core where the IPC will be acceptable. Their results showed big improvements in throughput and turnaround time when compared to the Linux scheduler, with negligible overheads and good fairness. Sondag and Rajans approach also involved no modification to the Linux scheduler.

To summarize, in the area of scheduling, there have been various attempts to solve scheduling problems using various forms of intelligent techniques [46-50]. We also discussed some approaches used to solve other problems, not CPU scheduling, but that have potential use in CPU scheduling.

### 6. SUMMARY OF RECENT TRENDS

Some researchers have the view that CPU scheduling is not a major problem because CPUs are very fast [6]. They prefer to focus on other components of the operating system, such as the I/O schedulers. Those researchers focusing on CPU scheduling tend to do low level work, making use of performance counters on the CPU to analyze behavior of various tasks on various processing elements. There are many opportunities for research in asymmetric multiprocessing, which researchers and the industry view as being very important for cutting costs and power consumption in the future. Table 1 shows the years and types of scheduling covered by our survey.

TABLE 1. Chronology of Various Data Mining Approaches to Solving Scheduling Problems Covered by Our Survey

| Type of Solution | Year | | | | | | | | | Total |
|---|---|---|---|---|---|---|---|---|---|---|
| | 1981 | 1993 | 1997 | 2001 | 2005 | 2006 | 2007 | 2008 | 2009 | |
| Distributed | - | - | - | - | - | - | 1 | - | - | 1 |
| Single Proc. | 1 | 1 | 1 | 1 | 2 | 2 | 3 | 4 | 1 | 16 |
| Multi Proc. | 1 | 1 | 1 | - | 1 | 1 | 2 | 1 | 1 | 9 |
| Asymmetric | - | - | - | - | - | - | - | 1 | 1 | 2 |
| Total | 2 | 2 | 2 | 1 | 3 | 3 | 6 | 6 | 3 | 28 |

## 7. CONCLUDING REMARKS

The approaches we have surveyed in this paper focus directly on CPU scheduling, or on systems, with potential for application in CPU scheduling. Definitely, as the technology matures, operating systems will be more autonomous, saving system administration time and costs, with parameters for various components, such as the scheduler, being set without human intervention. Various types of data mining methods will play a big part in relevant new technologies.